\documentclass[runningheads]{llncs}
\usepackage{graphicx}

\usepackage[misc,geometry]{ifsym}
\usepackage{booktabs}
\usepackage{amssymb}
\usepackage{multirow}

\usepackage{listings}
\usepackage{url}
\usepackage{pgf-pie}
\usepackage{booktabs}
\usepackage{dot2texi}
\usepackage{amssymb}
\usepackage{diagbox}
\usepackage{makecell}
\definecolor{logging-blue}{RGB}{1, 130, 172}
\definecolor{dark-red}{rgb}{0.3,0.1,0.1}
\definecolor{dark-green}{rgb}{0.1,0.3,0.1}
\definecolor{dark-blue}{rgb}{0.1,0.1,0.5}
\usepackage[colorlinks,linkcolor=dark-red,citecolor=dark-green,urlcolor=dark-blue]{hyperref}

\usepackage[ruled,lined,linesnumbered]{algorithm2e}
\DontPrintSemicolon
\SetKwFunction{LearningPhase}{learning\_phase}

\usetikzlibrary{shapes, arrows, patterns, automata, positioning}

\newcommand\copyrighttext{%
  \footnotesize \textcopyright~IFIP International Federation for Information Processing 2024. Published by Springer Nature International Publishing Switzerland 2024. All rights reserved. The final publication is available at Springer via \href{http://dx.doi.org/10.1007/978-3-031-56326-3\_21}{http://dx.doi.org/10.1007/978-3-031-56326-3\_21}. 
  
  Cite this article as follows: Sadlek, L., Čeleda, P. (2024). Cyber Key Terrain Identification Using Adjusted PageRank Centrality. In: Meyer, N., Grocholewska-Czuryło, A. (eds) ICT Systems Security and Privacy Protection. SEC 2023. IFIP Advances in Information and Communication Technology, vol 679. Springer, Cham. https://doi.org/10.1007/978-3-031-56326-3\_21}
\newcommand\copyrightnotice{%
\begin{tikzpicture}[remember picture,overlay]
\node[anchor=south,yshift=12pt] at (current page.south) {\fbox{\parbox{\dimexpr\textwidth-\fboxsep-\fboxrule\relax}{\copyrighttext}}};
\end{tikzpicture}%
}

\begin{document}
\title{Cyber Key Terrain Identification Using Adjusted PageRank Centrality}

\author{Luk\'{a}\v{s} Sadlek \inst{1,2} \Letter \orcidID{0000-0003-2577-6633} \and
Pavel \v{C}eleda\inst{1,2}\orcidID{0000-0002-3338-2856}}

\authorrunning{Luk\'{a}\v{s} Sadlek, Pavel \v{C}eleda}

\institute{Institute of Computer Science, Masaryk University, Brno, Czech Republic \and
Faculty of Informatics, Masaryk University, Brno, Czech Republic
\email{sadlek@mail.muni.cz, celeda@fi.muni.cz}\\}
\maketitle              
\copyrightnotice
\begin{abstract}
The cyber terrain contains devices, network services, cyber personas, and other network entities involved in network operations. Designing a method that automatically identifies key network entities to network operations is challenging.
However, such a method is essential for determining which cyber assets should the cyber defense focus on.
In this paper, we propose an approach for the classification of IP addresses belonging to cyber key terrain according to their network position using the PageRank centrality computation adjusted by machine learning. 
We used hill climbing and random walk algorithms to distinguish PageRank's damping factors based on source and destination ports captured in IP flows. The one-time learning phase on a static data sample allows near-real-time stream-based classification of key hosts from IP flow data in operational conditions without maintaining a complete network graph.
We evaluated the approach on a dataset from a cyber defense exercise and on data from the campus network. The results show that cyber key terrain identification using the adjusted computation of centrality is more precise than its original version. 

\end{abstract}

\section{Introduction}
The current need in cybersecurity is to interpret security as a business risk, according to Gartner~\cite{gartner}. Therefore, organizations focus on achieving the resilience of their missions that define objectives fulfilled by services and supporting assets, e.g., technologies~\cite{certrmm}. Cybersecurity entities essential for mission execution are classified as \textit{cyber key terrain} from the perspective of the organization's operations~\cite{guion2017cyber}. However, research often deals with a not specified mission in practice, when an organization requires only to maintain its operations, and these entities are called \textit{critical assets}. An example of a critical asset from an implicit mission is a local domain name server (DNS) that allows for locating network resources.

Identification of \textit{cyber key terrain} determines network devices, network services, cyber personas, and other network entities that provide an advantage for attackers and defenders~\cite{guion2017cyber}. It is used for security assessment of cyber assets~\cite{barreto2019cyber}, probabilistic mission impact assessment~\cite{motzek2017context,sun2015}, mission-centric risk assessment~\cite{silva2018}, and mission-centric decision support~\cite{musman2011}. 
It is of utmost importance for achieving security because cyber defense can become meaningless without knowing which cyber assets to protect.  
Solutions for asset discovery can determine the types of devices (e.g.,~\cite{orionplatform}) and provide an asset inventory~\cite{netboxdoc}, but they often require tagging the critical devices manually.
Therefore, automated approaches that determine critical cyber assets from raw network monitoring data can verify the content of populated asset inventory and its tags of critical devices.

An indispensable position in the asset criticality classification belongs to 
network centrality measures~\cite{kim2011determining}, identifying the most influential network entities according to the position in a graph, i.e., centrality. For example, the PageRank measure by \textit{Brin and Page}~\cite{brin1998} estimates the importance of web pages based on references by other web pages and the importance of these referencing web pages. Researchers also claim that the most critical network paths contain assets with high centrality~\cite{stergiopoulos2015using}, and high vertex centrality indicates asset criticality~\cite{kay2021identification}.
However, these methods can only be used to determine criticality plausibly if they consider computer network specifics, i.e., differences among vertices and edges from the network~\cite{kim2011determining}.

In this paper, we deal with two research questions. The first is: \textit{How to determine which IP addresses from cyber terrain are the key according to the network communication?} The second question is related to the evaluation of the approach: \textit{Does adjusting the PageRank centrality lead to better correctness of determining the cyber key terrain, and can it process IP flows from the real-world network?} Our focus is on the essential properties of IP flows~\cite{rick2014flow}, i.e., IP addresses, ports, and timestamps. 

We contributed to the current state by studying PageRank-based cyber key terrain identification adjusted by optimization methods called hill climbing~\cite{jacobson2004analyzing} and random walk~\cite{selman1994noise}. These machine learning methods used static data to learn different damping factors for different port pairs to optimize output PageRank values. The learning success was measured using the harmonic mean of the criticality classification's precision and recall (F1 score). Estimated damping factors were then used for dynamic stream-based computation of PageRank centrality on IP flow data. The approach was evaluated using a public dataset from a cyber defense exercise~\cite{tovarnak2020data} and a campus network.

This paper is organized as follows. Section \ref{sec:sota} describes the current state of cyber key terrain mapping, network centrality measures, hill climbing and random walk methods, and IP flow. Section \ref{sec:method} proposes the approach for the identification of key IP addresses using the adjusted PageRank centrality. Consequently, Section \ref{sec:evaluation} provides the evaluation and discusses the results, their importance, and the method's limitations. Last, Section \ref{sec:conclusion} concludes the paper.

\section{Related Work} \label{sec:sota}
The related work consists of four parts. The first part describes cyber key terrain mapping in general. The second part explains how centrality measures are used to estimate the criticality of network entities. The third part is dedicated to hill climbing and random walk methods. The last part describes IP flow, its types, and its properties.

\subsection{Cyber Key Terrain Mapping}
Cyber key terrain mapping identifies entities essential for attackers and defenders from five cyberspace planes -- the supervisory, the cyber persona, the logical, the physical, and the geographic plane~\cite{raymond2014}. For example, the logical plane, which is the most relevant plane for this paper, describes the software, network services, IP addresses, and domain names. 
Usual mapping approaches are based on crown jewels analysis, impact dependency graphs, and ontologies~\cite{guion2017cyber}. The crown jewels analysis determines cyber assets (i.e., crown jewels) that perform mission-critical functions~\cite{musman2011}, access other crown jewels, protect them, or enable them to work correctly. 
The impact dependency graphs contain assets, services, missions, mission steps, and their dependencies~\cite{guion2017cyber}.
Ontologies represent the mission domain using data entities, their properties, and their relationships.

Mentioned approaches use various data sources to identify cyber terrain. 
For example, \textit{Goodall et al.}~\cite{Goodall2009} used LDAP queries, NetFlow traffic, Unix logs, and FTP logs. \textit{Sun et al.}~\cite{sun2015} used system call logs containing operations on files, processes, and sockets. \textit{Motzek and Möller}~\cite{motzek2017context} analyzed captured network traffic using Wireshark. As a result, created mission models contain various entities. 
\textit{Musman et al.}~\cite{musman2011} applied business process modeling during crown jewels analysis to express the mission using its activities and necessary IT assets. Cyber-ARGUS~\cite{barreto2019cyber} uses an impact graph depicting dependencies among tasks, services, and cyber assets. A dependency graph by \textit{Motzek and Möller}~\cite{motzek2017context} contains business resources, functions, processes, and companies as vertices. \textit{Silva et al.}~\cite{silva2018} created a metamodel expressing mission, business processes, network services, infrastructure nodes, software, and their relationships.

The dynamic nature of cyber key terrain complicates its identification,
which requires aging out old data related to mission-critical components \cite{noel2018mission} and modeling information that flows between them~\cite{guion2017cyber}.
In our opinion, research works often focus on high-level models while not considering all the necessary details of filling the model with information extracted from raw data. 
Besides, manual modeling of missions is infeasible for large networks. These aspects must be addressed by methods that identify critical network assets.

\subsection{Network Centrality Measures}
Centrality measures form a significant group of methods for determining asset criticality~\cite{kim2011determining} based on position in a graph.
The most used variants are degree, betweenness, closeness, and eigenvector centralities. Degree centrality uses degrees of vertices, and the betweenness assigns higher criticality to vertices located between clusters, i.e., bridges. Closeness favors vertices close to others due to the easy transmission of resources in the network. Finally, eigenvector centrality considers the importance of neighbors connected by edges~\cite{kim2011determining}. The PageRank centrality belongs to eigenvector centralities but redistributes the vertex importance only among vertices linked by its outgoing edges.

PageRank centrality can be computed iteratively based on equation:
\begin{equation} \label{eq:basic}
PR_v = \frac{1-d}{n} + d \cdot \sum_{(u,v)\in E} \frac{PR_u}{deg^{out}(u)}
\end{equation} 
where $u$ and $v$ denote vertices, $n$ is the count of vertices in a graph, $E$ denotes a set of edges, and $d$ is a damping factor. $PR_v$ and $PR_u$ are the PageRank centralities of vertex $v$ in the current iteration and vertex $u$ from the previous iteration, and $deg^{out}(u)$ is equal to the number of outgoing edges from vertex $u$. The damping factor with a default value of $0.85$ represents the probability that the random surfer on web page $v$ will continue with a random web page~\cite{brin1998}. 

The most relevant variant of the original PageRank algorithm for our paper is Temporal PageRank~\cite{rozenshtein2016}, processing a stream of timestamped edges representing interactions between vertices. It uses the same damping factor as the static PageRank. However, it adds another transition probability that influences how fast the temporal PageRank converges to the static one because it describes the probability that the random surfer will choose the next edge from the stream to continue the random walk that follows links between vertices~\cite{rozenshtein2016}. 

Centrality measures cannot be applied to network data directly if they consider all vertices and relationships equal~\cite{kim2011determining}. Therefore, researchers often adjust the input graph for centrality measures. For example, \textit{Stergiopoulos et al.}~\cite{stergiopoulos2015using} used dependency risk paths consisting of asset dependencies for risk mitigation. Another centrality-based method was proposed by \textit{Kay et al.}~\cite{kay2021identification} to identify critical assets using a PageRank-based criticality score in the infrastructure network representing resource supplies between assets. 
On the contrary, \textit{Oliva et al.}~\cite{oliva2020aggregating} applied different perspectives from the degree, betweenness, eigenvector, and PageRank centrality measures to identify the most influential vertices in a network. However, in our opinion, it is worth adjusting the computation of centrality measures instead of adjusting their input or combining them. 

\subsection{Hill Climbing and Random Walk}
Hill climbing and random walk are machine learning methods used for finding optimal solutions to optimization problems, e.g., problems related to the boolean satisfiability problem (SAT). In general, the hill climbing method is usable for local search of solution space consisting of considered variables with respect to an objective function that defines the best solution. Its iterative algorithm 
slightly modifies the current solution to achieve a better neighboring solution~\cite{jacobson2004analyzing}, which often differs in the value of one variable. 

The hill climbing algorithm can find a local optimum different from the global one. Therefore, it is often combined with the random restart, which assigns random values to all variables after finding the local optimum~\cite{jacobson2004analyzing}. However, we can also add noise to the search algorithm using the random walk method instead of waiting until the local optimum is reached~\cite{selman1994noise}. The random walk method differs in assigning a random value to one conflict variable with some probability. For example, the conflict variable in the SAT problem is the one that appears in an unsatisfied clause~\cite{selman1994noise}. Other approaches from this area include, e.g., simulated annealing and modified versions of the mentioned methods.

\subsection{IP Flow}
The paper focuses on two types of IP flows -- unidirectional and bidirectional. A unidirectional IP flow is a set of IP packets with common properties (e.g., source and destination IP address, source and destination transport port) observed by the network observation point during a specific time window~\cite{rick2014flow}. On the contrary, the bidirectional flow contains packets sent between two endpoints in both directions. The source of bidirectional flow is determined by the initiator of the first observed packet, by network perimeter, or arbitrarily~\cite{trammell2008bidirectional}. 

Many applications using IP flows require their ordering. The start timestamp of IP flow in the IP Flow Information eXport (IPFIX) protocol is determined according to the timestamp of the first observed packet from the flow. In contrast, the end timestamp is the timestamp of the flow's last packet or packet before timeout will cause the flow to be exported~\cite{rick2014flow}.
 
\section{Method for Cyber Key Terrain Identification} \label{sec:method}
Our method for cyber key terrain identification at the logical cyberspace layer determines key IP addresses necessary for network operations. The innovation of the PageRank centrality measure consists of considering communication-specific damping factors based on IP flows. We use hill climbing and random walk methods to learn the damping factors that numerically encode critical IP addresses inside a previously unknown network environment. Further, the computation phase quickly processes IP flow data using a stream-based PageRank version.

\subsection{Learning Phase}
The learning phase is based on the iterative PageRank algorithm. The previous Equation~\ref{eq:basic} implies that all PageRank values sum to one during one iteration when all PageRank values from the previous iteration summed to one because  
\begin{equation}
    \sum_{v \in V} {PR_v} = (1 - d) + d \cdot \sum_{v \in V} {PR'_v} = 1 - d + d = 1
\end{equation}
where $PR'_v$ denotes the PageRank value of the vertex $v$ in the previous iteration that appears exactly $deg^{out}(u)$ times in the sum. We modified Equation~\ref{eq:basic} to consider the specifics of network communication by using damping factors adjusted to source and destination port pairs. We obtained
\begin{equation} \label{eq:adjusted}
    PR_v = \frac{1}{n} - \frac{PR'_v}{deg^{out}(v)} \cdot \sum_{w \in out(v)} {d_{vw}} + \sum_{u \in in(v)} {\left(d_{uv} \cdot \frac{PR'_u}{deg^{out}(u)}\right)}
\end{equation}
where $d_{uv}$ denotes the damping factor for edge $(u, v)$ and $out(v)$ denotes the set of vertices $w$ such that edge $(v, w)$ exists in the graph. Similarly, $in(v)$ denotes a set of vertices from which an edge leads to the vertex $v$. Finally, the sum of all $n$ PageRank values using adjusted damping factors is
\begin{equation}
    \sum_{v \in V} {PR_v} = n \cdot \frac{1}{n} + \sum_{(v,u) \in E} {\left(d_{vu} \cdot \frac{PR'_v}{deg^{out}(v)} - d_{vu} \cdot \frac{PR'_v}{deg^{out}(v)}\right)} = 1
\end{equation}
because $d_{vu} \cdot \frac{PR'_v}{deg^{out}(v)}$ appears in the sum with positive and negative signs.

Iterative Algorithm \ref{algo:learning} uses the hill climbing and the random walk methods to estimate the best values of damping factors. It starts with preprocessing the input graph containing port pairs that appear in more than some small fraction of IP flows, e.g., 0.1\% (see also Section \ref{sec:evaluation}). Each node is assigned the initial PageRank value of $\frac{1}{n}$ and other necessary attributes, e.g., its IP address and predecessors with their port pairs. Initial damping factors for all processed source and destination port pairs are equal to default values of $0.85$. 

\begin{algorithm}[h]
\SetKwFunction{Preprocessing}{preprocessing}
\SetKwFunction{OneIterationOfPagerank}{one\_iteration\_of\_pagerank}
\SetKwFunction{AssignBestScore}{assign\_best\_F1\_score}
\SetKwFunction{ChooseConflictNode}{choose\_random\_conflict\_node\_and\_its\_port\_pair}
\SetKwFunction{HillClimbing}{hill\_climbing}
\SetKwInOut{Input}{Input}\SetKwInOut{Output}{Output}

\Input{graph, max\_iterations, probability, results, heuristic}
\Output{best F1 score, best damping factors}
\BlankLine
    preprocessing()\;
    one\_iteration\_of\_pagerank(graph, factors, results)\;\label{alg:first_iteration}
    iterations $\gets$ 0\;
    \While{F1\_score $\neq$ 1 {\bf and} iterations $\leq$ max\_iterations}{
        assign\_best\_F1\_score()\;\label{alg:f1_score}
        port\_pair $\gets$ choose\_random\_conflict\_port\_pair()\;
        \If{random\_experiment $>$ $1$ $-$ probability}{
            factors[port\_pair] $\gets$ random(0, 1)\label{alg:random_walk}}
        \lElse{hill\_climbing(heuristic)\label{alg:hill_climbing}}
        iterations $\gets$ iterations + 1\;
        one\_iteration\_of\_pagerank(graph, factors, results)\;
    }
    assign\_best\_F1\_score()\;

\caption{Learning phase} \label{algo:learning}
\end{algorithm}

The first PageRank iteration on line~\ref{alg:first_iteration} processes the current damping factors for port pairs and updates centrality values for all nodes using Equation~\ref{eq:adjusted}. It returns the F1 score and the list of misclassified nodes based on the ground-truth labels for individual IP addresses. Nodes are considered critical if their centrality is above the threshold of $\frac{1}{n}$. The F1 score is a suitable measure of classification's correctness for imbalanced datasets where most IP addresses are noncritical.

The while condition checks whether the computation is completed, i.e., all nodes are classified correctly or the maximum number of iterations is achieved. Then we update the best F1 score on line~\ref{alg:f1_score} if necessary and randomly choose a port pair located on the edge leading to a misclassified (conflict) node. We continue with a random walk or hill climbing based on the probability from a random experiment. The solution space of optimization consists of damping factors and their values, while the objective function is the F1 score. 

The random walk on line~\ref{alg:random_walk} chooses a random damping factor for the current port pair from 0 to 1 and allows escaping from the local minimum found by hill climbing. On the contrary, the hill climbing on line~\ref{alg:hill_climbing} gradually assigns values from 0 to 1 by 0.05 steps for the damping factor to find one that improves the F1 score for PageRank iteration. We use four heuristics to decide whether to change a damping factor for the port pair, if it has the same F1 score as before hill climbing. The first one, the \textit{maximum} heuristic, always rewrites the current damping factor and achieves the maximum of changes. On the contrary, the \textit{minimum} heuristic never rewrites the current value and achieves the minimum of changes. The third one uses an \textit{average} of consecutive allowable values. Last, \textit{the smallest difference} heuristic uses default values of $0.85$ when possible to differ as least as possible from the default PageRank factors. When the loop is executed, the highest F1 score achieved during all iterations and damping factors valid during the best iteration form the result.

\subsection{Computation Phase}
Learned values of damping factors for individual port pairs are then used in the dynamic streaming PageRank computation using~\cite{rozenshtein2016} since a computer network can produce such a large amount of data that it cannot be processed as a static graph. We adjusted the streaming algorithm by replacing each appearance of one common damping factor with a specific damping factor $d_{uv}$ valid for the edge $(u, v)$. We also assume that one edge corresponds to one IP flow. The algorithm passes through the list of edges only once and has linear memory complexity with respect to the count of vertices. All port pairs not present during the learning phase are assigned the default damping factors equal to $0.85$.

The stream-based algorithm allows near real-time processing of IP flows in practice with respect to time constraints. It is expected that the PageRank values for individual nodes may vary throughout the computation because the dynamic data causes some noncritical devices to obtain high centrality for a short time. Therefore, we advise focusing on high centrality in a longer time window because the approximate algorithm needs to converge to the static PageRank.

\section{Evaluation} \label{sec:evaluation}
The method was evaluated on data captured in a network infrastructure emulated during a cyber defense exercise~\cite{tovarnak2020data} and in the campus network of Masaryk University. In both cases, we used 70\% of the data for learning damping factors on static graphs. The static graphs contained IP addresses and port pairs with a number of occurrences above 0.5\% of total IP flows for data from the cyber defense exercise and 0.1\% for university network data. These graphs also did not contain duplicate edges between IP addresses and timestamps. Consequently, the whole datasets were used to evaluate the stream-based PageRank variant.

We used 1,000 iterations of the while loop from Algorithm~\ref{algo:learning}, which should provide enough time to improve damping factors demonstrably. The probability of applying random walk was set to $0.1$, which implies that computation usually applied hill climbing. Moreover, we accomplished several runs for each heuristic. Averaged results are compared with the performance of the PageRank algorithm with the default damping factors of $0.85$. 

The stream-based variant used transition probability $\beta=0.5$, expressing that the random surfer can follow the next edge with equal probability as ignoring it in the random walk. Since PageRank values of nodes dynamically change during stream-based computation, we regularly evaluated samples after some count of IP flows. The evaluation was accomplished on a personal computer with 16 GB RAM, four CPU cores, and a processor's clock speed of 3.3 GHz. The method was implemented in Python. 

\subsection{Dataset from Cyber Defense Exercise}
The first IP flow dataset contains data captured during a two-day cyber defense exercise in 2019 and ordered according to the observation time~\cite{tovarnak2020data,tovarnakzenodo}, which also corresponds to sorting by the end timestamp. The network topology consisted of six equal networks (blue teams 1 -- 6) and one global network providing services for all teams. Despite being provided with the same network topology, all teams behaved differently during the exercise, which allows for evaluating the method on six partial datasets. The flow capture interface captured bidirectional flows and was located between the networks of the teams and the global network. We consider only the forward direction of bidirectional flow from the initiator of the connection 
because it corresponds to the dependency the centrality measure should consider, and the backward direction contains a reply.

\begin{table}[t]
    \centering
    \caption{F1 scores for heuristics in the learning phase according to the networks of individual teams from the cyber defense exercise (T1 -- T6). F1 scores for ten-minute-long (U10m) and one-hour-long capture from the university network (U1h).}
    \begin{tabular}{l|c|c|c|c|c|c|c|c}
        \toprule
        \textbf{Heuristic} & \textbf{T1} & \textbf{T2} & \textbf{T3} & \textbf{T4} & \textbf{T5} & \textbf{T6} & \textbf{U10m} & \textbf{U1h}\\ \midrule
        Default PageRank & 0.03 & 0.02 & 0.04 & 0.04 & 0.05 & 0.09 & 0.65 & 0.47\\
        Minimum & 0.75 & 0.77 & 0.72 & 0.74 & 0.77 & 0.60 & 0.84 & 0.75\\
        Maximum & 0.65 & 0.72 & 0.64 & 0.67 & 0.74 & 0.58 & 0.84 & 0.74\\
        Average & 0.66 & 0.67 & 0.65 & 0.65 & 0.76 & 0.58 & 0.83 & 0.75\\
        The smallest difference & 0.71 & 0.77 & 0.69 & 0.71 & 0.76 & 0.59 & 0.83 & 0.74\\ \midrule
    \end{tabular}
    \label{tab:learning_f1}
\end{table}

\begin{table}[t]
    \centering
    \caption{The size of the processed graph for learning and measured time. Results are divided according to the networks of individual teams.}
    \begin{tabular}{l|l|r|r|r|r|r|r}
        \toprule
         & & \textbf{T1} & \textbf{T2} & \textbf{T3} & \textbf{T4} & \textbf{T5} & \textbf{T6}\\ \midrule
        \multirow{3}{0.35cm}{\rotatebox[origin=c]{90}{\textbf{Data}}} & Nodes & 554 & 1,380 & 542 & 884 & 503 & 219\\
        & Edges & 1,468 & 3,064 & 1,631 & 2,418 & 1,361 & 584\\ 
        & IP flows & 66,499 & 116,897 & 63,400 & 88,734 & 78,254 & 30,781\\ \midrule
        \multirow{3}{*}{\rotatebox[origin=c]{90}{\textbf{Time}}} & Preprocessing & 0.47 s & 1.35 s & 0.44 s & 0.91 s & 0.39 s & 0.09 s\\ 
        & Learning time & 175.04 s & \textbf{17.49 min} & 168.33 s & 7.41 min & 142.85 s & 30.20 s\\ 
        & Computation time & 1.34 s & 2.06 s & 1.22 s & 1.81 s & 1.56 s & 0.58 s\\ \bottomrule
    \end{tabular}
    \label{tab:learning_exercise}
\end{table}


We labeled critical hosts from team networks and the global network manually. Table~\ref{tab:learning_f1} contains averaged best F1 scores of learning on the static graphs for individual heuristics, while the default PageRank contains a converged value for comparison. The adjusted method was much better during the whole phase due to estimating better damping factors for criticality classification. Results from Table~\ref{tab:learning_exercise} also show that the method's execution time depends on the graph size in the learning phase and the total number of IP flows. Both phases are feasible in practice because the learning phase will be run only once, or we can combine factors learned on several partial graphs. The method will require learning the factors again if the probability distribution of the most frequent IP addresses and port pairs will considerably change.

\begin{figure}[t!]
  \centering
  \includegraphics[width=.495\textwidth]{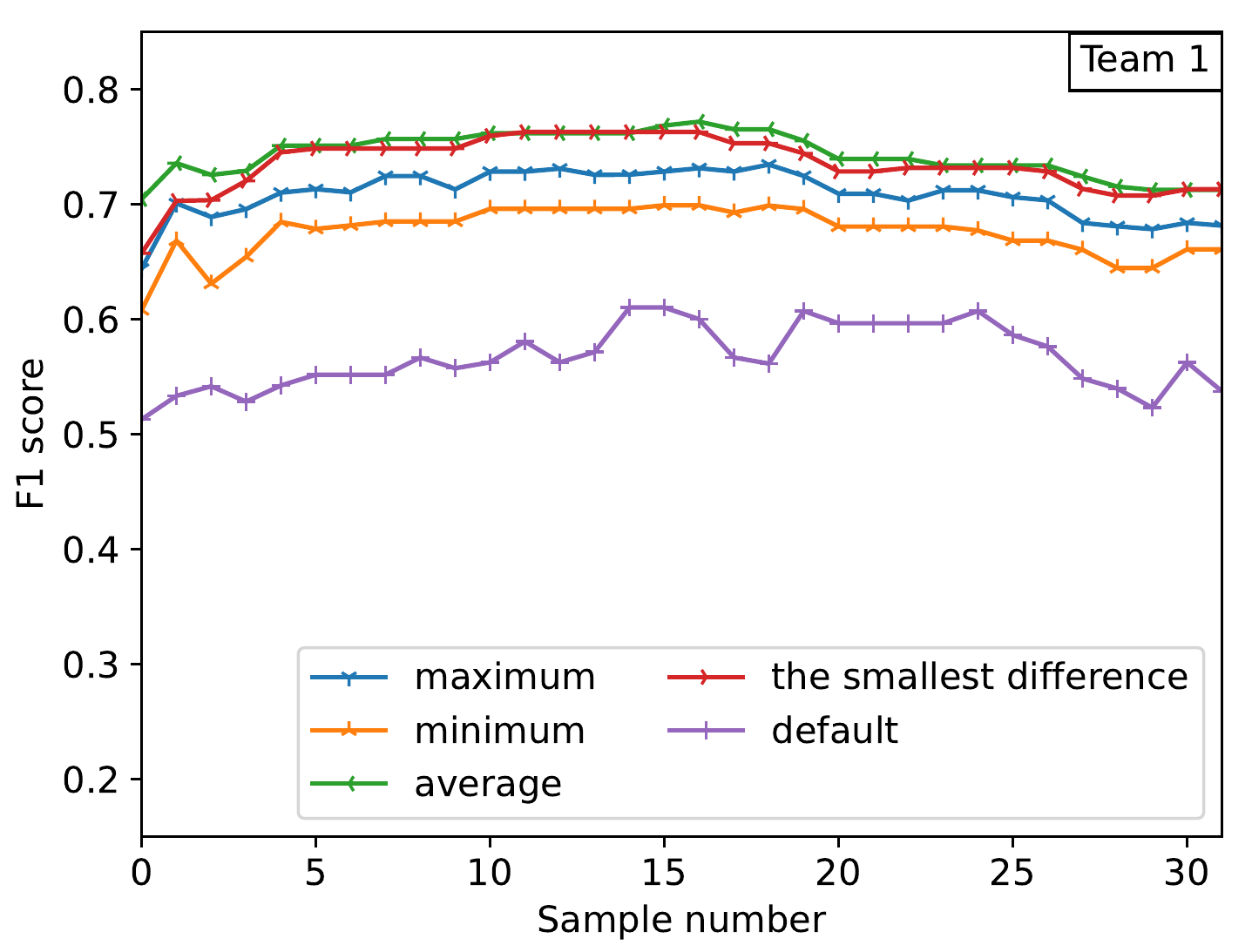}
  \includegraphics[width=.495\textwidth]{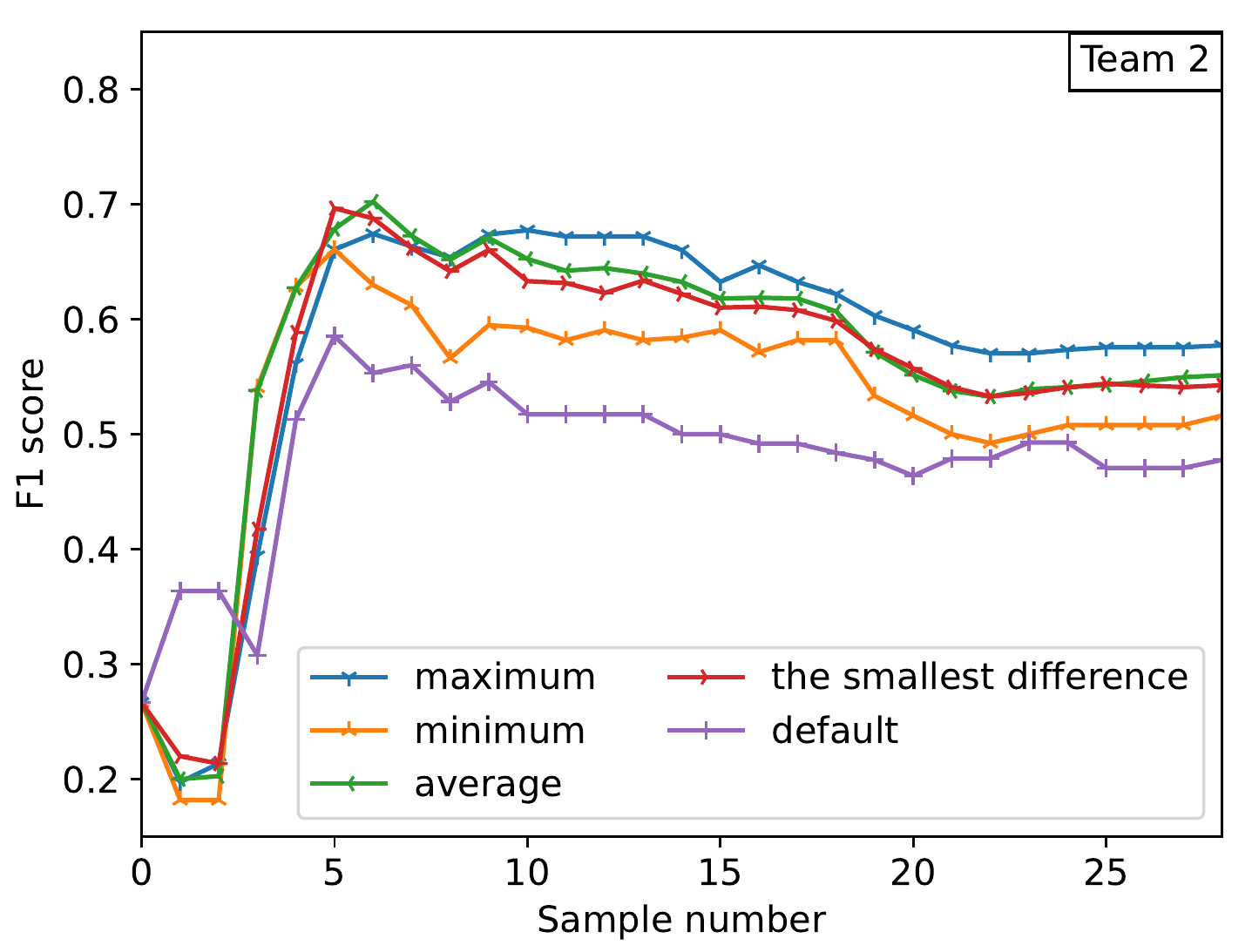}

  \includegraphics[width=.495\textwidth]{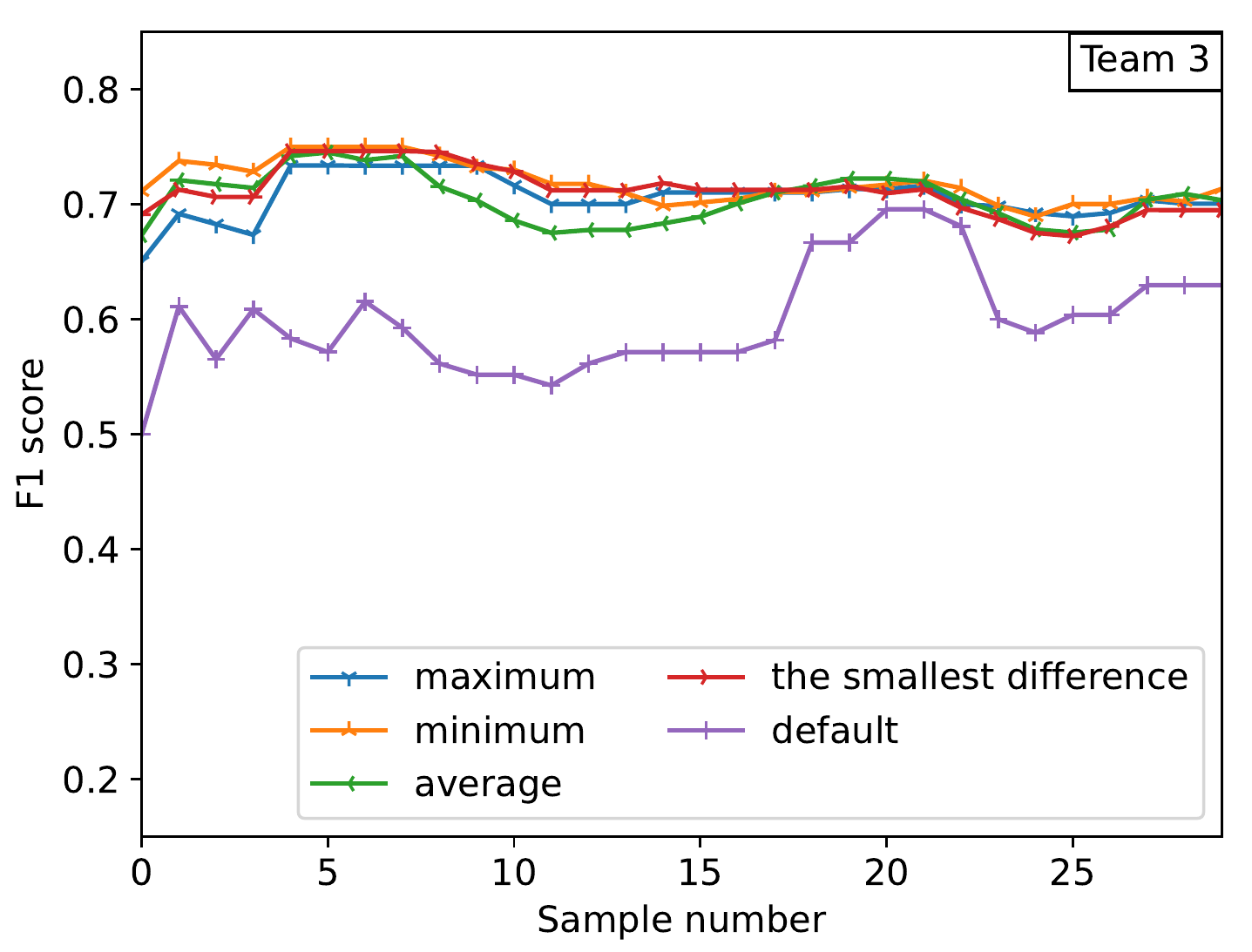}
  \includegraphics[width=.495\textwidth]{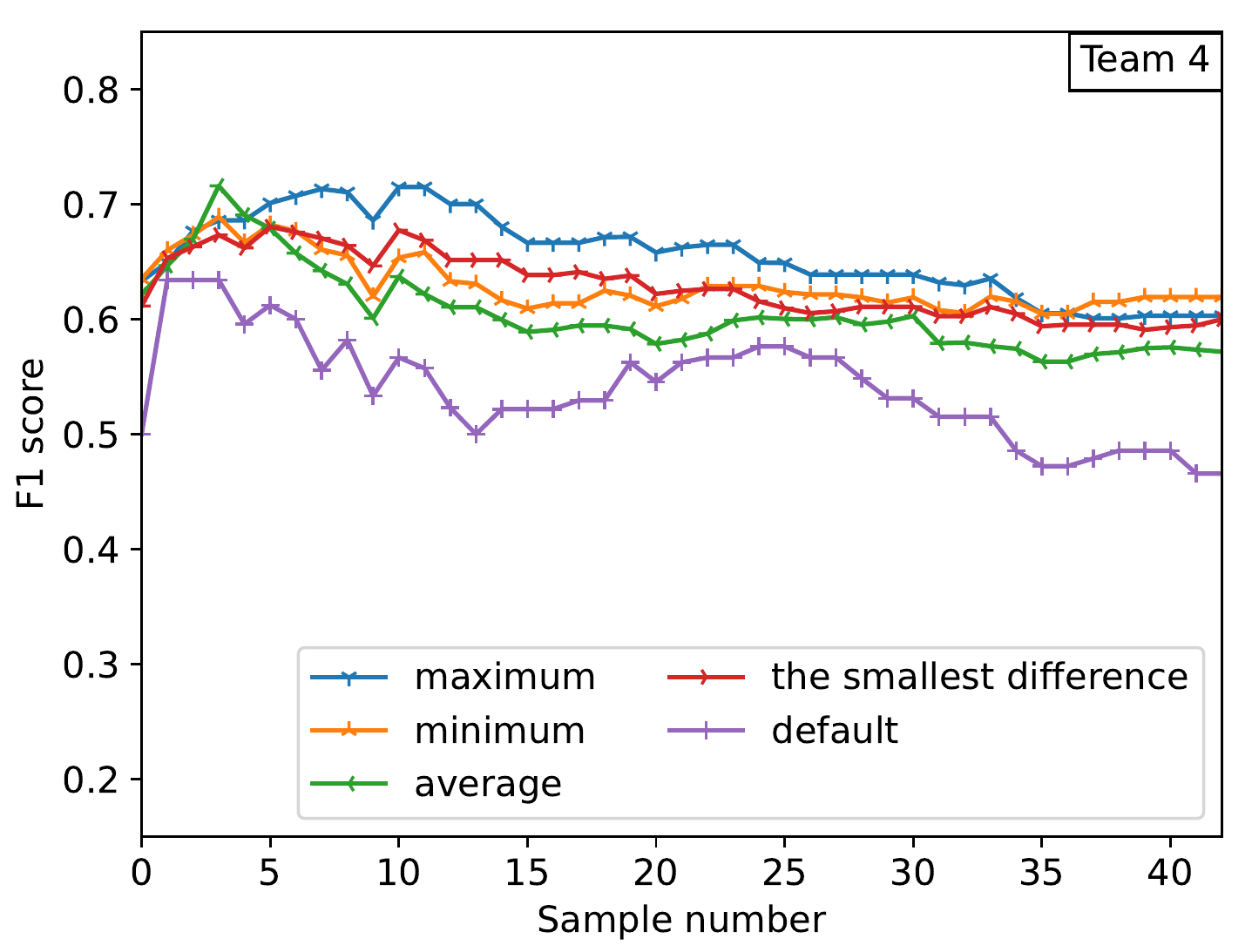}

  \includegraphics[width=.495\textwidth]{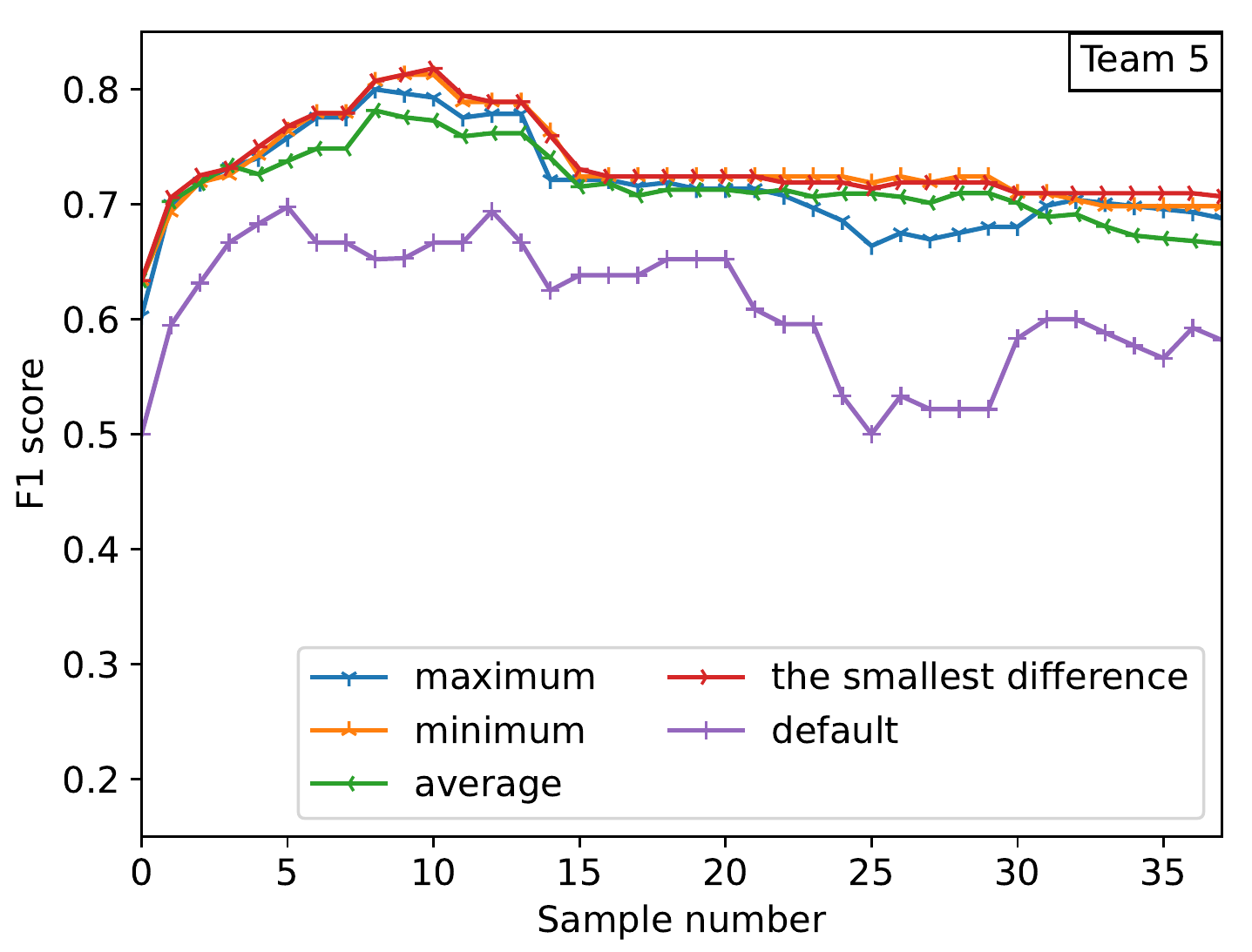}
  \includegraphics[width=.495\textwidth]{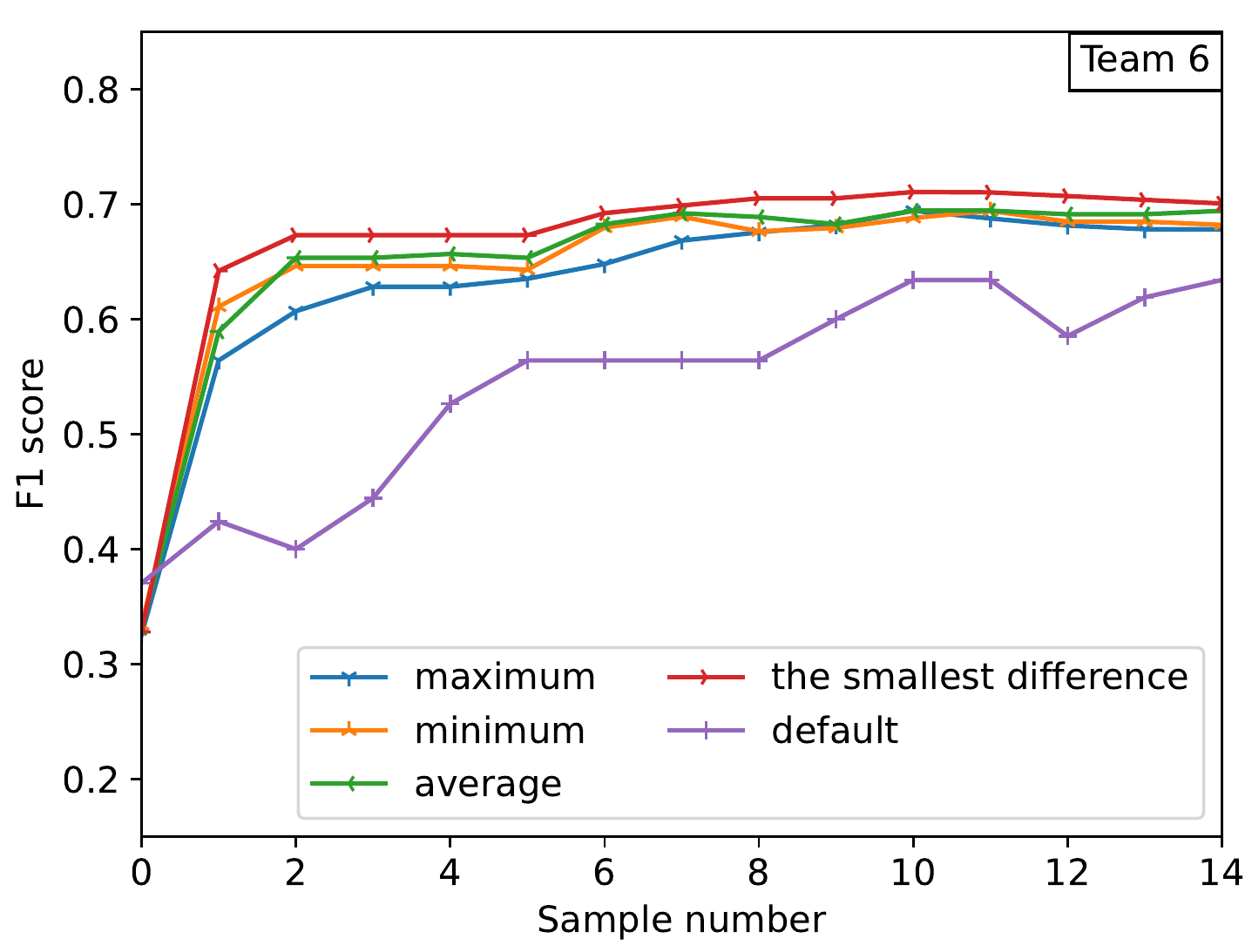}

  \caption{Line graphs containing F1 scores for heuristics and the PageRank with default damping factor divided according to six team networks.} \label{fig:cyber_czech}
\end{figure}

Figure~\ref{fig:cyber_czech} contains average F1 scores for the stream-based computation phase obtained from five measurements per each heuristic, i.e., 20 measurements for each team. In each iteration, we used newly learned damping factors. We observed the classification's progress after 4,000 loaded IP flows for the second team and 2,000 IP flows for other teams. In general, all heuristics provide almost always better F1 scores than the default PageRank, focusing only on source and destination ports, and follow a similar pattern. The comparison should be considered from a long-term perspective due to fluctuating activity of hosts. 

\subsection{Dataset from the Campus Network}
The university network is assigned class B address space with /16 CIDR prefix. Network probes in the campus network were situated at the network edge. In our evaluation, we used IP flows from a ten-minute window captured during working hours on one Tuesday in March 2022 and a one-hour window captured during working hours on one Wednesday in February 2023. We executed the method ten times for each heuristic and accomplished five or eight samples of results during each execution. Data from the ten-minute-long window were unordered, but we sorted them according to the start timestamps, while in the second case, we used ordering according to the end timestamps.

The learning phase used a static graph that contained 90 vertices and 1,569 edges (ten-minute capture) and 78 vertices and 3,605 edges (one-hour capture), which could be preprocessed in 0.18 and 0.58 seconds on average. F1 scores from the learning phase are listed in Table~\ref{tab:learning_f1}, where heuristics achieved better F1 scores compared to the PageRank with default damping factors. The hill climbing and random walk with 1,000 iterations took 15.25 and 28.92 seconds on average. Finally, stream-based processing of 8.56 million flows representing ten-minute-long IP flow capture took only 81.66 seconds on average (including sorting). On the contrary, the one-hour-long IP flow capture contained 89.28 million flows and was processed in less than nine minutes. Results show that the method can quickly process large amounts of IP flow data.
 
A hard problem was obtaining result labels since we could not explicitly enumerate all critical IP addresses. Therefore, we focused on an overview of network subnets maintained by network administrators. It contains an organization unit (e.g., faculty) and a short description for each subnet. Critical IP addresses belong to subnets that can contain a lot of critical devices, according to the description. However, the overview does not provide exact but consistent labels. After evaluating the results, we manually added false positives that were critical to the labels and re-executed the whole method in a way that did not devaluate the results of the default PageRank.

False positives confirmed the correctness of criticality classification. External false positives were usually Google servers (1e100.net), Facebook servers, domains for downloading updates to user devices, and generally important services, such as Google DNS. They are often influenced by the activity of students.
False positives from the internal network are typically devices from the wireless network with predetermined IP ranges, staff devices, VPN addresses, and hosts providing less essential network services. These false positives were often quickly replaced by another in the ten-minute-long dataset.

Table~\ref{tab:results_mu} shows that the minimum, the maximum, and the smallest difference heuristics increase the count of recommended critical devices and devices from the campus network in the sorted ten-minute-long time window. The most significant improvement was achieved by the minimum heuristic that recommends approximately two to five more critical devices compared to the default damping factors. The variance describes an opportunity to tune damping factors using heuristics in practice.
When executed on unordered data, the PageRank with default values achieved approximately the same results differing at most in only one device. The average heuristic achieved better results on non-sorted data, but the other heuristics did not.

\begin{table}[t]
    \centering
    \caption{The number of true positives in the top 100 results according to samples (S1 -- S5), the number of hosts from the university network (N1 -- N5), and the average variance of true positives during the ten-minute-long window.}
    \begin{tabular}{l|c|c|c|c|c||c|c|c|c|c||c}
        \toprule
        \textbf{Heuristic} & \textbf{S1} & \textbf{S2} & \textbf{S3} & \textbf{S4} & \textbf{S5} & \textbf{N1} & \textbf{N2} & \textbf{N3} & \textbf{N4} & \textbf{N5} & \textbf{Var} \\ \midrule 
        Default PageRank & 24 & 24 & 23 & 23 & 23 & 35 & 32 & 30 & 28 & 29 & --- \\
        Minimum & \textbf{26.4} & \textbf{28.3} & \textbf{27.6} & \textbf{27.9} & \textbf{27.8} & \textbf{41.4} & \textbf{37.8} & \textbf{37.8} & \textbf{37.0} & \textbf{35.7} & \textbf{18.9}\\
        Maximum & 25.2 & 25.8 & 24.5 & 24.2 & 24.2 & 38.1 & 34.7 & 33.9 & 32.4 & 31.4 & 13.6\\
        Average & 22.8 & 23.3 & 22.3 & 22.6 & 22.4 & 33.6 & 30.2 & 29.3 & 27.8 & 27.6 & 10.3\\
        The smallest difference & 24.0 & 24.5 & 23.5 & 23.4 & 23.4 & 36.9 & 33.4 & 32.6 & 30.8 & 30.2 & 7.72\\ \bottomrule
    \end{tabular}
    \label{tab:results_mu}
\end{table}
In the case of one-hour-long capture sorted according to the end timestamps, the top 100 IP addresses contained more critical IP addresses from the university network, even for default PageRank (see Table~\ref{tab:results_mu_hour}). Therefore, the improvement is not visible among the top 100 results because the top results form only a small part of all results. The best heuristics, in this case, are alternately maximum, minimum, and the smallest difference, while the average heuristic demonstrates the worst criticality classification. 

\begin{table}[t]
    \centering
    \caption{The number of true positives in the top 100 results according to samples (S1 -- S8), the average number of hosts from the university network (N), and the average variance of true positives in these samples during the one-hour-long window.}
    \begin{tabular}{l|c|c|c|c|c|c|c|c||c||c}
        \toprule
        \textbf{Heuristic} & \textbf{S1} & \textbf{S2} & \textbf{S3} & \textbf{S4} & \textbf{S5} & \textbf{S6} & \textbf{S7} & \textbf{S8} & \textbf{N} & \textbf{Var} \\ \midrule 
        Default PageRank & \textbf{41} & 41 & \textbf{43} & \textbf{42} & \textbf{41} & \textbf{41} & \textbf{41} & \textbf{41} & 59.1 & --- \\
        Minimum & 40.3 & 41.5 & 41.1 & 40.7 & 40.1 & 40.1 & 40.3 & 39.9 & \textbf{59.2} & 2.7 \\
        Maximum & 40.5 & \textbf{41.7} & 41.2 & 40.8 & 40.4 & 39.9 & 39.9 & 39.3 & 58.3 & \textbf{11.2} \\
        Average & 38.6 & 39.8 & 39.7 & 39.1 & 38.7 & 38.0 & 38.6 & 38.3 & 56.9 & 7.8\\
        The smallest difference & 40.4 & 41.4 & 41.1 & 40.6 & 40.5 & 40.3 & 40.2 & 39.9 & 58.4 & 3.8 \\ \bottomrule
    \end{tabular}
    \label{tab:results_mu_hour}
\end{table}

\subsection{Limitations} \label{sec:lessons}
Several possible limitations of the proposed method and the performed evaluation need to be discussed. First, samples of results may not be accomplished at the right moments, and necessary progress could remain hidden. 
Second, the static graph for learning can only contain the most influential vertices and port pairs. The learning phase should be accomplished on the same type of flows (unidirectional or bidirectional) to estimate consistent damping factors. Moreover, damping factors should be tuned to their best values in practice, and other IP flow attributes could be considered, e.g., length of communication.

Third, the streaming algorithm~\cite{rozenshtein2016} was designed for short interactions with only one timestamp. However, IP flows contain start and end timestamps and may not be optimally sorted, e.g., because of timeouts that will flush incompleted flows from a collector. Furthermore, the wrong forward direction of bidirectional flow, e.g., because of the not captured first packet, can also influence the results. 
Last, a large amount of IP flows will not fit into the main memory. However, maintaining a sliding window of IP addresses with the highest centrality and the total value of centrality for the removed vertices throughout the algorithm is feasible. This value would be equally divided among the retained IP addresses. 

\section{Conclusion} \label{sec:conclusion}
In this paper, we proposed a method for determining which IP addresses from cyber terrain are the key by adjusting PageRank centrality. We used two machine-learning methods -- hill climbing and random walk -- to distinguish damping factors according to the source and destination ports. Despite using only essential properties of IP flows, we showed that this approach leads to better correctness of classifying critical hosts compared to the default damping factors, except for the natural temporal fluctuation of the stream-based PageRank variant. The evaluation also proved that the PageRank centrality is suitable for determining IP addresses related to critical network's and organization's services.

When using only the top 100 results, the precision between using default and estimated damping factors was almost equal with the increased count of flows. The method can process IP flows from the real-world network using a stream-based PageRank algorithm with the estimated damping factors in practice.
Supplementary materials at~\cite{sadleksupplementary} contain a proof-of-concept implementation of the learning and computation phases with the ground-truth labels for IP addresses from the cyber defense exercise data published in~\cite{tovarnakzenodo}.

\section*{Acknowledgement}
This research was supported by ERDF project ``CyberSecurity, CyberCrime and Critical Information Infrastructures Center of Excellence'' (No. CZ.02.1.01/0.0/\\0.0/16\_019/0000822).

%

\bibliographystyle{splncs04}
\bibliography{bibliography}

\end{document}